\documentclass{aa}

\usepackage{txfonts}
\usepackage{natbib}
   \bibpunct{(}{)}{;}{a}{}{,}
\usepackage{epsfig}
\usepackage{graphicx}
   \DeclareGraphicsExtensions{.eps, .jpg}
\usepackage{color}
\newcommand{\Msun}{M_{\odot}}
\newcommand{\Mdot}{\dot{M}}
\newcommand{\MdotEdd}{\dot{M}_{\rm Edd}}
\newcommand{\LEdd}{L_{\rm Edd}}
                                                   
\begin{document}

%=============================================================
\title{
Investigating the mass of the intermediate mass black hole candidate HLX-1 with the {\tt slimbh} model
}
%=============================================================
\author
{
Odele Straub \inst{1} \thanks{E-mail: odele.straub@obspm.fr (OS)} 
\and
Olivier Godet \inst{2,3} 
\and 
Natalie Webb \inst{2,3}
\and
Mathieu Servillat \inst{4,1}
\and
Didier Barret \inst{2,3}
}
%=============================================================
\institute{
Laboratoire Univers et Th\'eories, CNRS UMR 8102, Observatoire de Paris, Universit\'e Paris Diderot, 5 Place Jules Janssen, F-92195 Meudon, France.
\and
Institut de Recherche en Astrophysique and Plan\'etologie (IRAP), Universit\'e de Toulouse, UPS, 9 Avenue du colonel Roche, F-31028 Toulouse Cedex 4, France
\and
CNRS, UMR5277, F-31028 Toulouse, France
\and
Laboratoire AIM, CEA Saclay, Bat. 709, F-91191 Gif-sur-Yvette, France
}
%=============================================================
\date{Draft version}
%=============================================================
\abstract{In this paper we present a comprehensive study of the mass of the intermediate mass black hole candidate HLX-1 in the galaxy ESO 243-49. We analyse the continuum X-ray spectra collected by {\it Swift}, {\it XMM-Newton}, and {\it Chandra} with the slim disc model, {\tt slimbh}, and estimate the black hole mass for the full range of inclination (inc = $0^{\circ} - 85^{\circ}$) and spin ($a_* = 0 - 0.998$). The relativistic {\tt slimbh} model is particularly suited to study high luminosity disc spectra as it incorporates the effects of advection, such as the shift of the inner disc edge towards smaller radii and the increasing height of the disc photosphere (including relativistic ray-tracing from its proper location rather than the mid-plane of the disc). We find for increasing values of inclination that a zero spin black hole has a mass range of 6,300 - 50,900 $\Msun$ and a maximally spinning black hole has a mass between 16,900 - 191,700 $\Msun$. This is consistent with previous estimates and reinforces the idea that HLX-1 contains an intermediate mass black hole. }
%=============================================================
\authorrunning{Straub et al.}
\titlerunning{Investigating the mass of HLX-1 with {\tt slimbh}}
%=============================================================
\keywords{accretion, accretion discs -- X-rays: binaries, black hole.}
\maketitle
%=============================================================

%=============================================================
\section{Introduction}
\label{sec:intro}
%=============================================================
%=============================================================
There is a limit to how luminous an object of a given mass can be. When a star or an accretion disc is in hydrostatic equilibrium it supports itself against gravity by its own internal radiation pressure. The critical luminosity (assuming isotropic emission) is thus given by the Eddington limit, $\LEdd = 4 \pi c \, G M / \kappa_{\rm es} = 1.26 \, \times \, 10^{38} \, (M/\Msun) \,\, {\rm erg/s}$, where c is the speed of light, G is the gravitational constant, M is the mass of the gravitating body, $\Msun$ is the solar mass, and $\kappa_{\rm es} = 0.2 \,(1+X) \, {\rm cm}^2/{\rm g}$ (with $X = 1$) is the electron scattering opacity of a pure hydrogen plasma. There are, however, objects whose luminosities exceed this natural limit. 

Ultraluminous X-ray sources (ULXs) are sources with X-ray luminosities $\gtrsim 10^{39} \, {\rm erg/s}$. Most ULXs are thought to be powered by super-Eddington accretion onto a stellar mass black hole which can be accomplished (i) by powering strong disc winds \citep{sha+73, lip99},  (ii) by advecting the radiation along with the flow as in radiation pressure dominated disc models like Polish doughnuts \citep{abr+78,jar+80} and slim discs \citep{abr+88}, or (iii) both, advection and outflows \citep{pou+07, dot+11}. Luminosities up to $10^{41} \, {\rm erg/s}$ can therefore still be explained by super-Eddington mass accretion rates onto stellar mass black holes which can have maximum masses up to $\sim 80 \, \Msun$. These higher mass black holes can be explained by direct collapse of metal poor stars \citep{bel+10}. 

The brightest known ULX in the sky is \object{2XMM J011028.1-460421} in the lenticular galaxy ESO 243-49 \citep[z = 0.0223, ][]{wie+10}. With peak luminosities $\sim 10^{42} \, {\rm erg/s}$ this object, dubbed HLX-1 \citep{far+09}, belongs to a subclass of ULXs called the hyper-luminous X-ray sources \citep[HLX,][]{gao+03}.  Like many X-ray binaries, HLX-1 shows transitions from low/hard to high/soft states \citep{god+09,ser+11}, transient radio emission that can be associated with hard-to-soft transitions \citep{web+12} and a weak optical counterpart \citep{sor+10}. The extremely high luminosity of HLX-1 suggests, however, the presence of an intermediate mass black hole (IMBH) with a mass of about $100 \Msun$ to $\sim 10^5 \Msun$. 
%=============================================================
% TABLE 1 -- DATA
%=============================================================
\begin{table*}[htp]
\centering
\caption{Observational data}
\begin{tabular}{ccccc}
\hline \hline 
Obs. Name & Instrument & Obs. ID & Start Date & End Date \\ 
\hline 
{\it Swift}        & XRT                  & 00031287(001-252) & 2008-Oct-24 & 2012-Sep-16 \\ [5pt]
                   &                      & 00032577(001-011) & 2012-Oct-02 & 2012-Nov-11 \\ [5pt]

{\it XMM - Newton} &  {\em pn},  \em MOS 1 $\&$ 2 & 0560180901        & 2008-Nov-28 & 2008-Nov-28  \\ [5pt]

{\it Chandra}      &  ACIS                & 13122             & 2010-Sep-06 & 2010-Sep-07 \\
\hline
\label{tab:data}
\end{tabular}
\end{table*}
%=============================================================
%=============================================================

Spectra of HLX-1 have already been studied with various disc models. These were either limited to one particular inclination and/or black hole spin, non- or semi-relativistic, or based on the standard \citet{sha+73} disc and therefore only valid in the lowest luminosity regime, $L \lesssim 0.1 \LEdd$. All previously used models agree that HLX-1 contains an IMBH. \citet{ser+11} predict a black hole mass $M > 9000 \Msun$ from fitting the non-relativistic {\tt diskbb} model to multi-epoch data collected by {\it Swift}, {\it XMM-Newton} and {\it Chandra}. \citet{dav+11} improve the mass constraints using a relativistic thin disc model with full radiative transfer, {\tt bhspec}, and find $3000 \Msun < M < 3 \times 10^5 \Msun$. They assume a mass range of $1778 - 316228 \Msun$, consider spins $a_* = -1$  to 0.99 and luminosities between $0.03 - 1 \LEdd$ and fit simultaneously for the degenerate mass and spin parameters. As a consequence, a large fraction of their fits peg at the boundary value of at least one of these free parameters, and the results from different spectra are inconsistent with each other. In addition, most of their fits require luminosities far higher than the standard disc models allows. \citet{god+12} address the last point by employing a simplified slim disc model that includes Comptonisation and some relativistic corrections \citep[][]{kaw03} and estimate $M \sim 2 \times 10^4 \Msun$. They fit for a large mass range between $1 - 10^5 \Msun$ and have a disc structure that allows for high luminosities, but their study is limited to a face-on disc around a Schwarzschild black hole. 

In this work we resolve the previous shortcomings regarding parameter space, luminosity regime and consistency among results from different spectra. We use a \emph{fully} relativistic slim disc model \citep{sad+11, str+11} that accounts on the one hand for effects related to high mass accretion rates such as advection of radiation, relocation of the inner disc edge towards radii smaller than the innermost stable circular orbit (ISCO), and extended disc height. On the other hand, the model incorporates full vertical radiative transfer by integrating the emission of local annuli spectra and ray-tracing from the proper photosphere location. This slim disc model, {\tt slimbh}, is valid for luminosities up to 1.25 $\LEdd$ and covers all inclinations and prograde spins (see the discussion of the model limits in Section~\ref{sec:modelling}). It differs from the well-known disc model {\tt bhspec}, which is based on a standard thin disc, only in the structure of the underlying disc \citep[the spectral differences between these two models have been studied in][e.g. their Section 3.2 and the top panel in their Fig.4]{str+13}. In comparison to the \citet{kaw03} slim disc, {\tt slimbh} is fully relativistic and uses full radiative transfer.  

We present results from fitting three thermal X-ray spectra of HLX-1 taken by {\it Swift}, {\it XMM-Newton}, and {\it Chandra}. With the employed slim disc model we are not only able to study the whole parameter plane spanned by black hole spin and inclination, we also get a consistent mass estimation for all spectra. This improves and solidifies the previous mass estimates. The paper contains a brief summary of the data selection in Section~\ref{sec:data}, a discussion about spectral fitting with the slim disc model in Section~\ref{sec:modelling}, and a summary of the results in Section~\ref{sec:results}. We finish with conclusions in Section~\ref{sec:discussion}

% ================================================
% ================================================
\section{X-ray data reduction}
\label{sec:data}
% ================================================
% ================================================
The slim disc model is, like other $\alpha$-disc models, based on a few specific assumptions that limit its applicability to thermal state spectra. In particular it assumes that each plasma annulus is locally in thermal equilibrium and optically thick so that it radiates like a blackbody. In order to obtain data of the highest possible statistical quality showing a clear soft excess that we can interpret as emission coming from a multicolour blackbody disc, we make a few amendments to the data selection used in previous studies by \citet{ser+11}, \citet{dav+11}, and \citet{god+12}. We summed up the {\it Swift} data around the outburst peak to improve the statistics and employ the latest calibration files for the {\it XMM-Newton} data from 2008 (see below), but we omitted the observation from 2004 where the source was most likely in a steep power-law state \citep{dav+11, god+12}. Only the {\it Chandra} spectrum remains the same. All data are listed in Table~\ref{tab:data}.

To enhance the visibility of a possible high energy tail the {\it Swift} spectrum consists of an accumulation of $\sim 121$\,ks of Photon Counting (PC) data near the outburst peak between 2009 and 2012 \citep[the plateau phase; see][]{god+12}. The {\it Swift}-XRT Photon Counting data (ObsID 31287 and 32577) were processed using the HEASOFT v6.14, the tool {\scriptsize XRTPIPELINE} v0.12.8\,\footnote{\url{http://heasarc.gsfc.nasa.gov/docs/swift/analysis/}}, and the calibration files ({\scriptsize CALDB} version 4.1). We used the grade 0-12 events, giving a slightly higher effective area at higher energies than the grade 0 events, and a 20 pixel (47.2 arcseconds) radius circle to extract the source and background spectra using {\scriptsize XSELECT} v2.4c. The background extraction region was chosen to be close to the source extraction region and in a region where we are sure that there are no sources present in the {\it XMM-Newton} field of view. The ancillary response files were created using {\scriptsize XRTMKARF} v0.6.0 and exposure maps generated by {\scriptsize XRTEXPOMAP} v0.2.7. We fit the spectrum using the response file {\scriptsize SWXPC0TO12S6$_{-}$20010101v012.RMF}. Each spectrum was grouped to contain a minimum of 20 counts per bin to optimise the $\chi^2$ technique.
% ================================================
% FIGURE -- 1
% ================================================
\begin{figure}[h]
\includegraphics[width=0.5\textwidth]{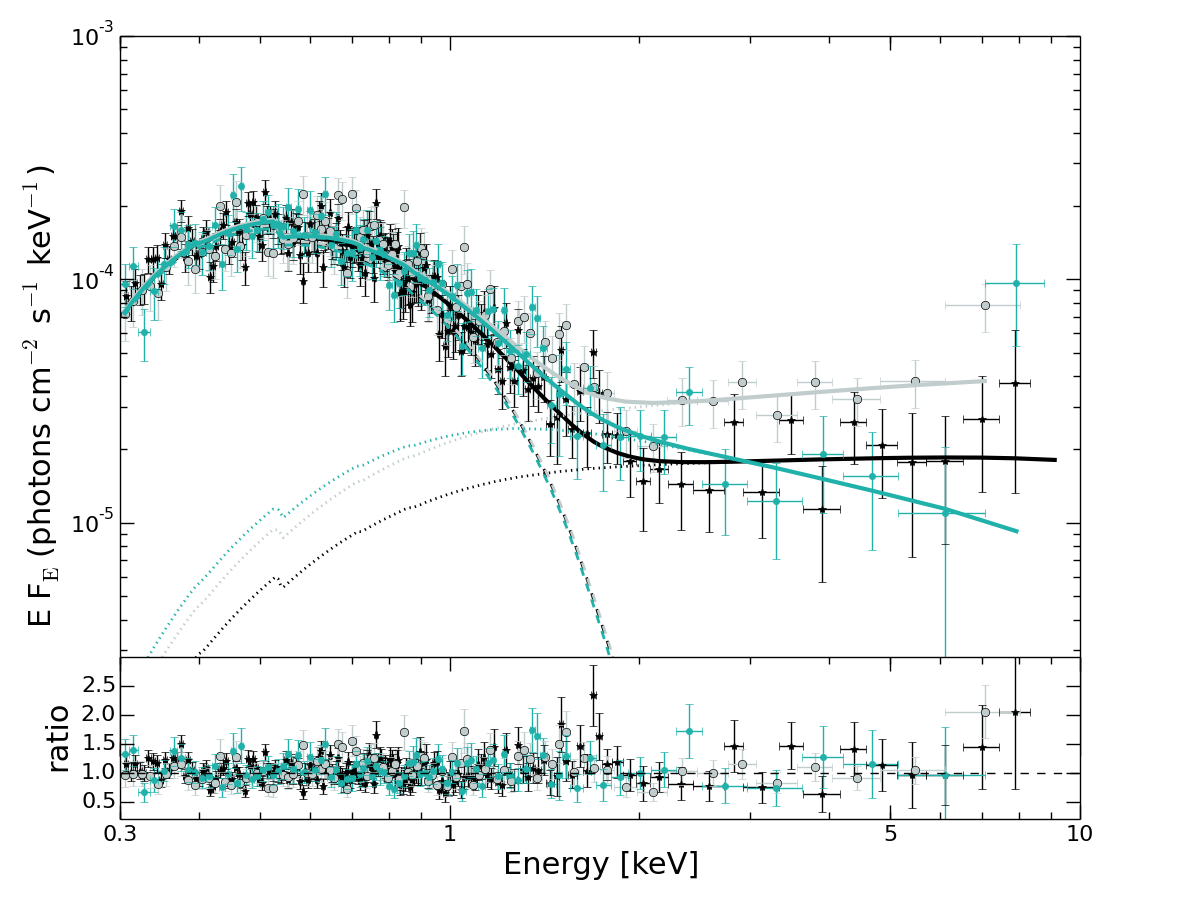}
\caption{
The {\it XMM-Newton} {\em pn} (black) and {\em MOS 1} and {\em MOS 2} (cyan and grey) data. Fits were performed with {\tt slimbh} for $a_* = 0.5$, $i = 60^{\circ}$ and $\alpha = 0.01$. The full model (solid lines) is composed of an absorbed slim disc component (dashed lines) and a Compton component (dotted lines). See Table~\ref{tab:results} for details.}
\label{fig:spectra}
\end{figure}
% ================================================

% ================================================
% FIGURE -- 2
% ================================================
\begin{figure*}[ht]
\includegraphics[width=\textwidth]{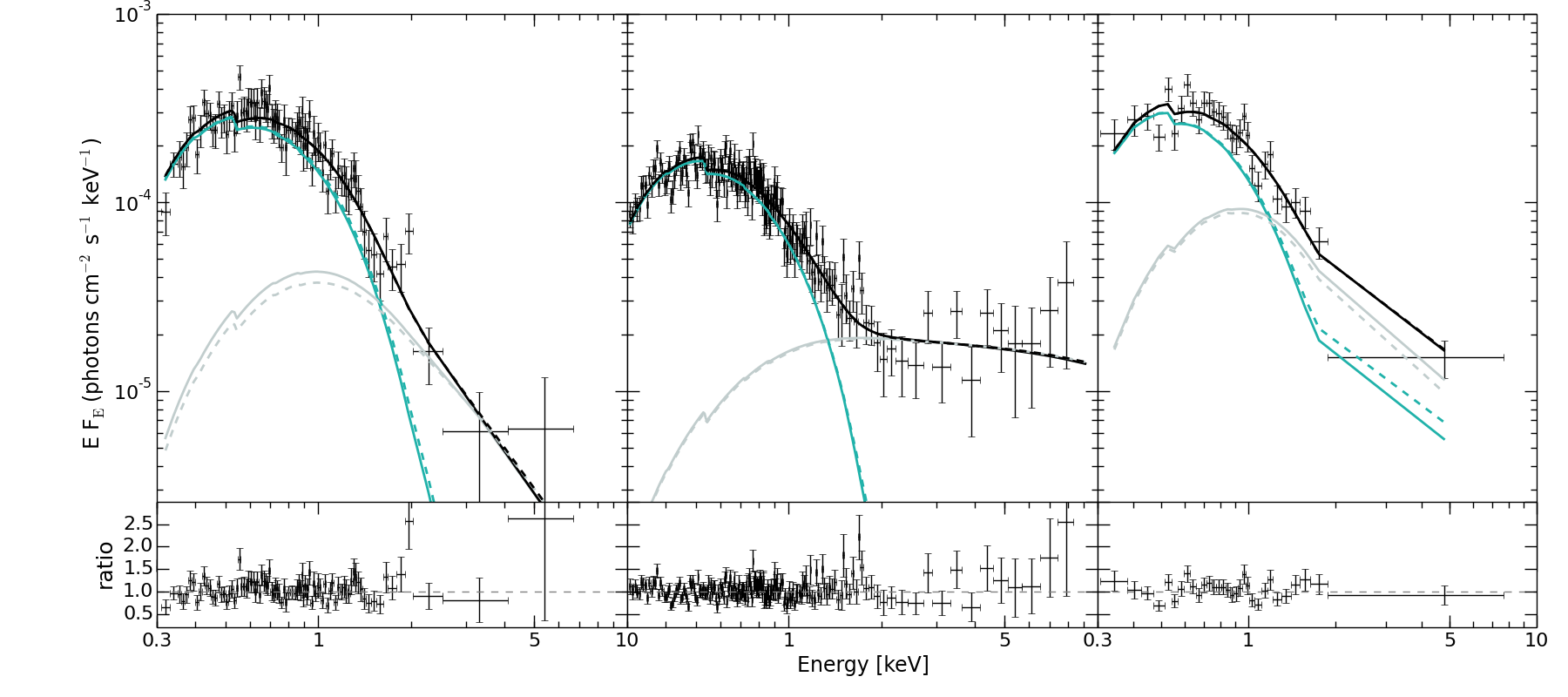}
\caption{
{\it Swift} (left), {\it XMM-Newton} {\em pn} (middle) and {\it Chandra} (right) spectra fitted with {\tt slimbh}. Each model consists of a disc and a Compton component (cyan and grey, respectively) and represents the best fit for the case $a_* = 0.5$, $i = 60^{\circ}$ and $\alpha = 0.01$ (compare Table~\ref{tab:results}). The dashed components indicate the best fits of the same configuration when $\alpha = 0.1$.}
\label{fig:spectra}
\end{figure*}
% ================================================

The {\em XMM-Newton} data were taken on 2008 November 28, whilst {\em HLX-1} was still in the high/soft state. To reduce the {\em XMM-Newton} data we used the latest version of the XMM-Newton Science Analysis Software (SAS, version 13.5) and the latest calibration files (CCFs, May 2014). The {\em MOS} (Metal Oxide Semi-conductor) data were taken in full frame mode using the thin filter and the data were reduced using the {\em emproc} task of SAS.  The event lists were filtered with the $\#XMMEA\_EM$ lag, and 0-12 of the predefined patterns (single, double, triple, and quadruple pixel events) were retained. The background was low and stable throughout the observation, resulting in 50.6 ks of clean data. We also filtered in energy, using the range 0.2-10.0 keV. The {\em pn-CCD} data (from the positive-negative Charge Coupled Device camera) were taken in the small window mode and using the thin filter. The data were reduced using the SAS {\em epproc} task and 0-4 of the predefined patterns (single and double events) were retained, as these have the best energy calibration.  Again the background was low and stable, resulting in 50.5 ks of clean data. We used the $\#XMMEA\_EP$ filtering and the same energy range as for the {\em MOS}. We extracted the data using the optimised source extraction region. The {\em MOS 1} spectra were extracted using circular regions with radii $36\arcsec$ and $43.5\arcsec$ for the source and background regions respectively. The background was chosen from a source free region close to the source. For the {\em MOS 2} we used regions of $33\arcsec$ and $43.5\arcsec$  respectively and for the {\em pn} we used $24\arcsec$ and $28\arcsec$ for the source and background. We rebinned the data into 5eV bins as recommended in the SAS threads\footnote{http://xmm.esac.esa.int/sas/current/documentation/threads/}. We used the SAS tasks {\em rmfgen} and {\em arfgen} to generate a {\em redistribution matrix file} and an {\em ancillary response file}, for each spectrum. The data were binned to contain at least 20 counts per bin.

% ================================================
% ================================================
\section{Modelling the thermal spectra of HLX-1}
\label{sec:modelling}
% ================================================
% ================================================

% ================================================
% TABLE 2 -- RESULTS
% ================================================
\begin{table*}[ht]
\centering
\caption{
Detailed results for {\it Swift}, {\it XMM-Newton} and {\it Chandra} spectra for $a_* = 0.5$ and $i = 60^{\circ}$. The two error intervals for the black hole mass corresponds to the 90\% and 3 $\sigma$ confidence level, respectively. The unabsorbed disc flux, $F_{\rm disc}$, is taken over the energy range 0.3 - 10 keV.
}
\begin{tabular}{c|ccc}
\hline \hline \\ [1pt]
Model Parameter & {\it Swift} & {\it XMM-Newton}  ({\em pn}, {\em MOS 1}, {\em MOS 2}) & {\it Chandra} \\ [5pt]
\hline \\ [1pt]
{\tt tbabs} & & & \\ [5pt]
%%%%%%%%%%%%%%%%%%%%%%%%%%
$N_H$ [$\times 10^{20}$]  & \multicolumn{3}{c}{4} \\ [10pt]
{\tt slimbh} & & & \\ [5pt]
%%%%%%%%%%%%%%%%%%%%%%%%%%
$M$ [$\Msun$] & $19989^{+5883 \; +13902}_{-3078 \;\; -4606}$ & $18473^{+1195 \; +2323}_{-1074 \; -1905}$ & $29952^{+14476 \; +36285}_{\;\; -8381 \; -12201}$ \\ [5pt]
$a_*$ & \multicolumn{3}{c}{0.5} \\ [5pt]
$L_{disc}$ [$\LEdd$] & $0.64^{+0.14}_{-0.19}$ & $0.43^{+0.02}_{-0.02}$, $0.43^{+0.03}_{-0.03}$, $0.42^{+0.03}_{-0.03}$ & $0.62^{+0.25}_{-0.18}$ \\ [5pt]
$i$ [deg] & \multicolumn{3}{c}{60} \\ [5pt]
$\alpha$ & \multicolumn{3}{c}{0.01} \\ [5pt]
D [Mpc] & \multicolumn{3}{c}{95} \\ [10pt]
{\tt nthc} & & & \\ [5pt]
%%%%%%%%%%%%%%%%%%%%%%%%%
$\Gamma$ & $4.16^{+3.24}_{-2.57}$ & $1.94^{+0.30}_{-0.25}$, $1.84^{+0.24}_{-0.21}$, $2.49^{+0.53}_{-0.43}$ & $4.55^{+2.58}_{-2.50}$ \\ [5pt]
$kT_e$ [keV] & \multicolumn{3}{c}{5} \\ [5pt]
$kT_{bb}$ [keV] & \multicolumn{3}{c}{0.2} \\ [5pt]
$z$ & \multicolumn{3}{c}{0.0223} \\ [5pt]
$N$ [$10^{-5}$] & $4.59^{+7.27}_{-4.23}$ & $1.45^{+0.45}_{-0.37}$, $2.38^{+0.61}_{-0.53}$, $2.52^{+0.90}_{-0.76}$ & $11.96^{+9.12}_{-10.36}$ \\ [10pt]
{\tt  pileup} & & & \\ [5pt]
%%%%%%%%%%%%%%%%%%%%%%%%%%
$t_{frame}$ [s] & - & - & 0.83 \\ [5pt]
$f$ & - & - & $0.16^{+0.61}_{-0.16}$ \\ [5pt]
\hline\\ [1pt]
$F_{disc}$ [$10^{-13} \, {\rm ergs/cm}^2/$s]  & 6.867 & 3.806 & 9.088 \\ [5pt]
\hline\\ [1pt]
$\chi^2$/dof ($\chi^2_r$) & 82.07/85 (0.97) & 367.34/361 (1.02) & 37.84/27(1.40) \\ [5pt]
\hline
\label{tab:results}
\end{tabular}
\end{table*}
% ================================================
% ================================================

The spectral model {\tt slimbh} \citep[after][]{abr+88, sad+11} is an additive disc model to be used in XSPEC \citep{arn96} and publicly available\footnote{\url{http://astro.cas.cz/slimbh}}. Like {\tt bhspec}, this is a relativistic $\alpha$-disc model with full radiative transfer but instead of using the standard thin disc it is based on a \emph{slim} disc, i.e. it incorporates the effects of advection. This means that with rising mass accretion rate an increasing fraction of photons gets trapped in the flow, carried inward, and is partly released at smaller radii. A typical {\tt slimbh} disc is therefore softer at photon energies below the spectral peak and harder above it in comparison to {\tt bhspec}.

The {\tt slimbh} model has nine parameters: Black hole mass $M$, black hole spin $a_*$, disc luminosity $L_{disc}$, inclination $i$, viscosity $\alpha$, distance $D$, hardening factor $f_{hard}$, limb darkening $lflag$ and vertical extent $vflag$. The latter two are flags to switch on/off the effect of limb darkening, and ray-tracing from the proper disc photosphere (both are switched on here). The hardening factor is calculated internally based on the TLUSTY grid of local annuli spectra \citep{hub+95} which are then integrated over the whole disc (alternatively, $f_{hard}$ could also be set to a constant colour correction factor). Using the cosmological parameters from the WMAP5 results ($H_0 = 71$ km s$^{-1}$ Mpc$^{-1}$, $\Omega_M = 0.27$ \& $\Omega_\Lambda=0.73$), we adopt a source distance of D = 95 Mpc. The observed luminosity temperature relation in X-ray binaries seems to favour low viscosities \citep{don+08}; we therefore fix the viscosity parameter at $\alpha = 0.01$. A discussion of how higher values of $\alpha$ affect the spectra can be found in Section~\ref{sec:viscosity}. The inclination is fixed at values between the model limits $0^{\circ} - 85^{\circ}$ in steps of $10^{\circ}$. The disc luminosity is left free to assume the best-fit value within the model limits $L_{disc} = 0.05 - 1.25 \LEdd$. The TLUSTY grid calculates the specific intensity in a given direction from plane-parallel layers of radiation, whereas the surface of a slim disc becomes slanted with respect to the equator when the luminosity increases. Our current upper limits on the luminosity and inclination parameter are set by hand and represent the limits up to which we are confident that the internal calculation of the hardening factor is accurate.

As in the {\tt bhspec} and {\tt kerrbb} models the black hole mass and spin are also in {\tt slimbh} degenerate parameters and cannot be fitted at the same time. When we leave both parameters free, the joint confidence contours for $M$ and $a_*$ tend to reveal either chains of local minima, a completely unconstrained $a_*$, or the pegging of the spin and/or luminosity parameter at a boundary value. To perform better fits and avoid such inconclusive results either the mass or spin parameter must be fixed. We thus freeze the spin at values between $a_* = 0.0 - 0.998$ in steps of 0.1 and fit for the mass. This procedure leads to well-constrained masses and luminosities that are consistent among the employed data sets. All three spectra show a notable amount of statistically low significance residuals below 2 keV that might be due to narrow emission lines \citep[see][]{god+12}. We do not try to fit these residuals and accept that the fit statistics for the {\it XMM} and {\it Chandra} spectra is slightly above the optimum.

The {\it Swift} and {\it XMM-Newton} data were fitted with the same model, {\tt tbabs} $\times$ ({\tt slimbh} + {\tt nthc}) and the {\it Chandra} data were in addition multiplied by {\tt pileup} \citep[see][]{ser+11}. The {\em pn}, {\em MOS 1}, and {\em MOS 2} data sets were fitted jointly with a tied mass parameter. The absorption component, {\tt tbabs}, is used to account for the total neutral hydrogen column density along the observer's line of sight, $N_H = 4 \times 10^{20} cm^{-2}$ \citep{far+09, god+12}. The thermal Compton component, {\tt nthc}, consists of the free power-law photon index $\Gamma$, the electron temperature $kT_e$, the seed photon temperature $kT_{bb} = 0.2$ keV \citep[][]{ser+11, god+12}, the input type for disc-blackbody seed photons, the redshift $z = 0.0223$ \citep{wie+10} and the normalisation (free). Because of the relatively large uncertainty in the data points at the highest energies the fits are not sensitive to the electron temperature. Its precise value neither influences the fitting parameters nor the goodness of fit. We therefore fix $kT_e = 5$ keV at an arbitrary value. The {\tt pileup} component uses a frame time of 0.83\,s and has only one free (grade morphing) parameter that can assume values between 0 and 1.  Figure~\ref{fig:spectra} and Table~\ref{tab:results} show the results of a fit with black hole spin $a_* = 0.5$ and inclination $i = 60^{\circ}$.

% ================================================
% ================================================
\section{The mass of HLX-1}
\label{sec:results}
\begin{figure*}[ht]
\includegraphics[width=0.5\textwidth]{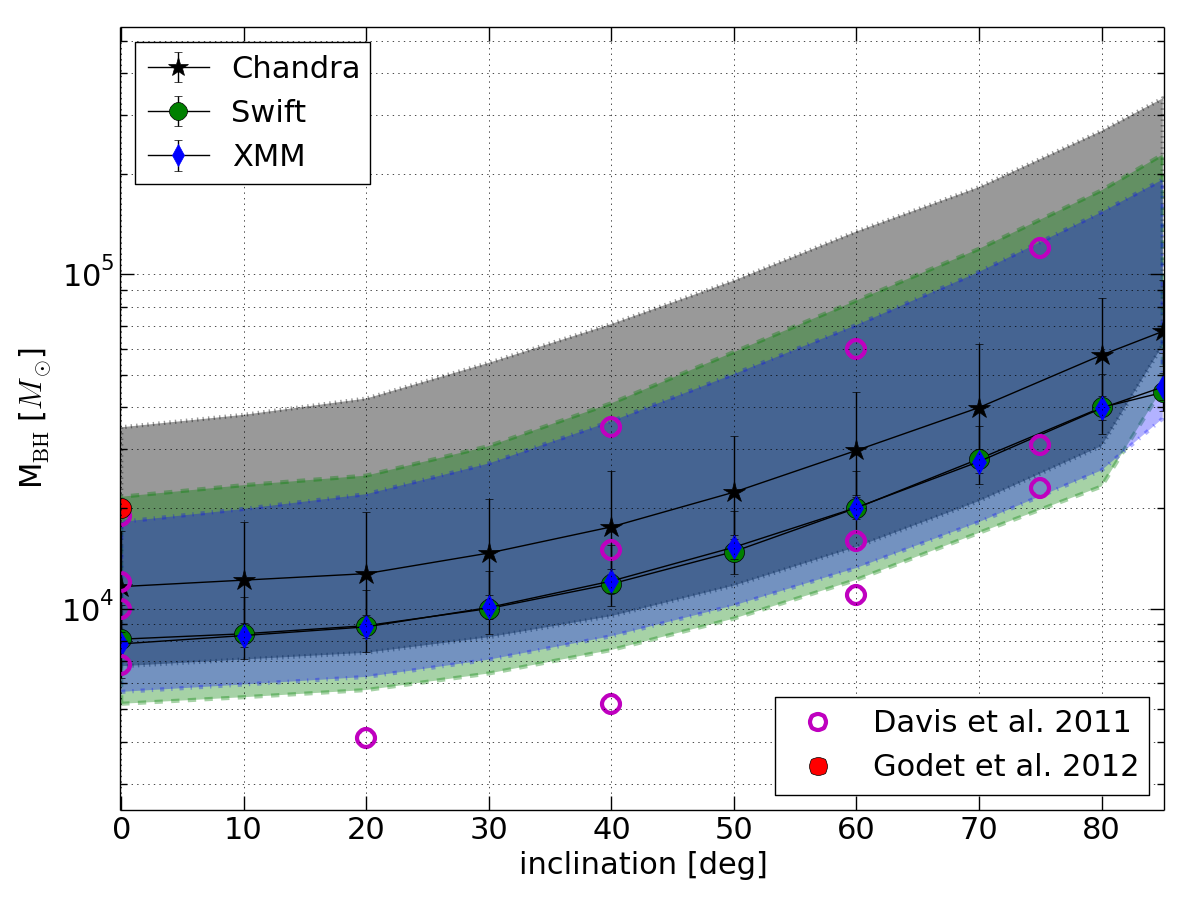}
\includegraphics[width=0.5\textwidth]{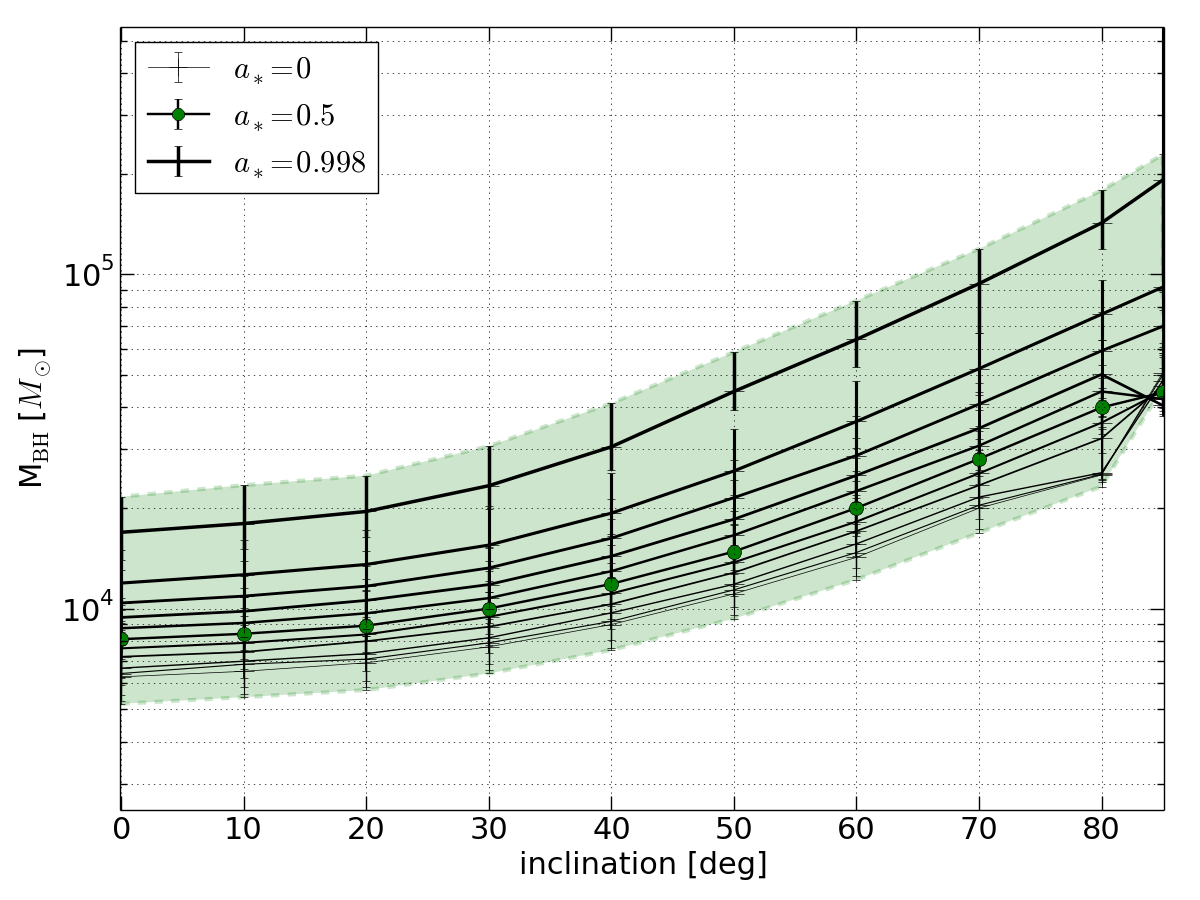}
\caption{The black hole mass in HLX-1. {\it Left}: The shaded areas indicate the 90\% confidence interval for the mass estimation of the {\it Chandra} (grey), {\it Swift} (green), and {\it XMM} (blue) spectrum, respectively, given for the whole spin range. The three solid lines represent the black hole mass that corresponds to the spin $a_* = 0.5$ (where the error bars denote 90\% confidence). The filled and open circles show previous mass measurements (see text for details). {\it Right}: The influence of spin on the mass estimation is shown for the {\it Swift} data. Error bars give the 90\% confidence interval.}
\label{fig:mass}
\end{figure*}

% ================================================
% ================================================
We study the parameter plane spanned by black hole spin and inclination and find the distribution of black hole mass shown on the left panel in Figure~\ref{fig:mass}. The shaded areas indicate the 90\% confidence interval for the whole spin range. Singled out as black lines are the masses (90\% confidence) we obtained for $a_* = 0.5$. The open and filled circles mark previous results (error bars have been omitted) of \citet{dav+11} and \citet{god+12}, respectively. The mass estimates from the {\it Chandra} spectrum (grey), which is affected by pileup, seem to slightly overestimate the black hole mass as compared to the {\it Swift} (green) and {\it XMM} (blue) spectra which are almost congruent. They are, however, consistent within the error bars. This behaviour does not apply to the results of \citet{dav+11}, where the highest mass at a given inclination is associated with a {\it Swift} or {\it XMM} spectrum and a spin that pegs at the maximum value. In general, higher inclinations implicate larger black hole masses as the geometrically induced reduced photon count is compensated with the larger emitting area of a higher source mass. With increasing spin the radiation originates from an ever smaller area around the black hole, which also translates into a reduced amount of photons and consequently a larger black hole mass. To show the effect of the black hole spin in detail we plot in the right panel of Figure~\ref{fig:mass} the masses for {\it Swift} spectrum alone. Given a Schwarzschild black hole and a face-on disc, the mass found by \citet{god+12} is substantially higher than the one found with models that include full radiative transfer.
% ================================================
% FIGURE -- 3
% ================================================
% ================================================

\begin{figure*}[ht]
 \includegraphics[width=0.5\textwidth]{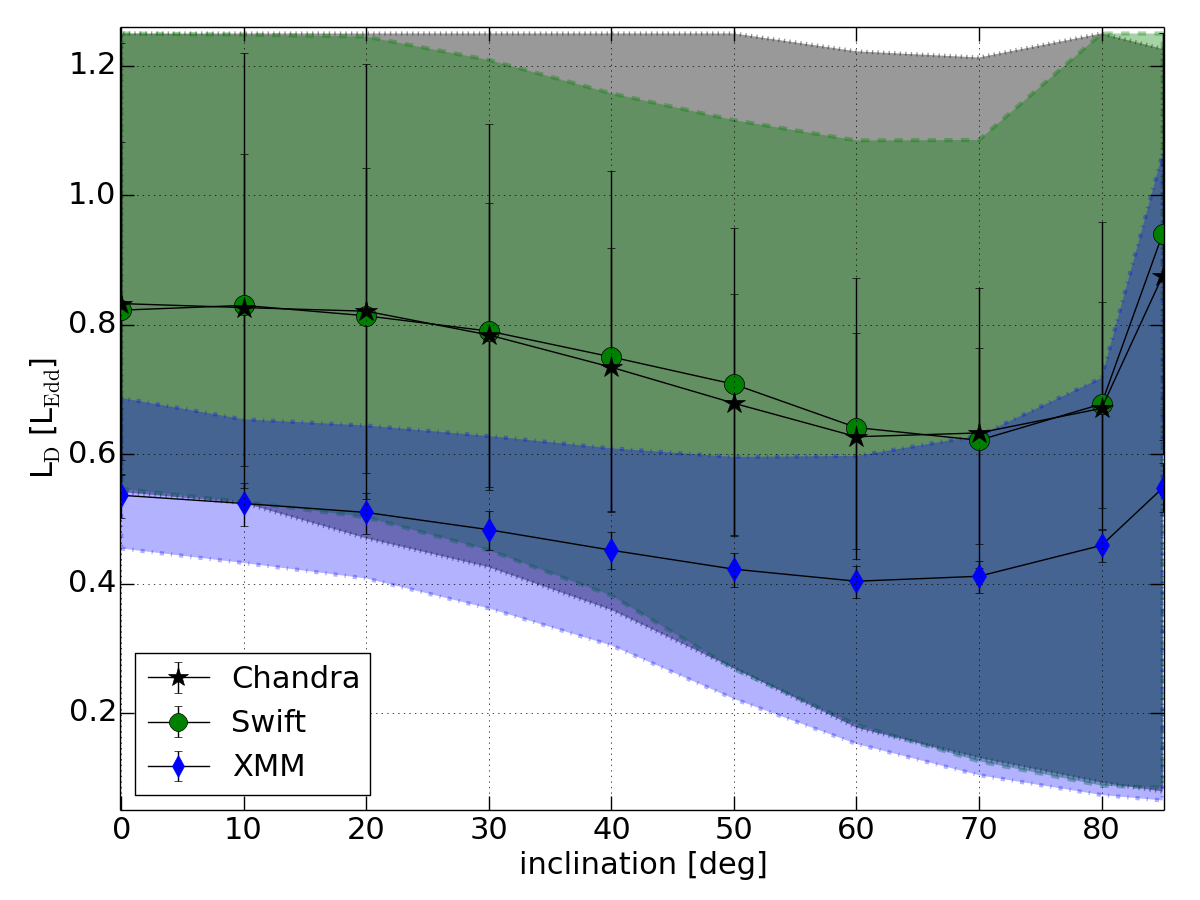}
 \includegraphics[width=0.5\textwidth]{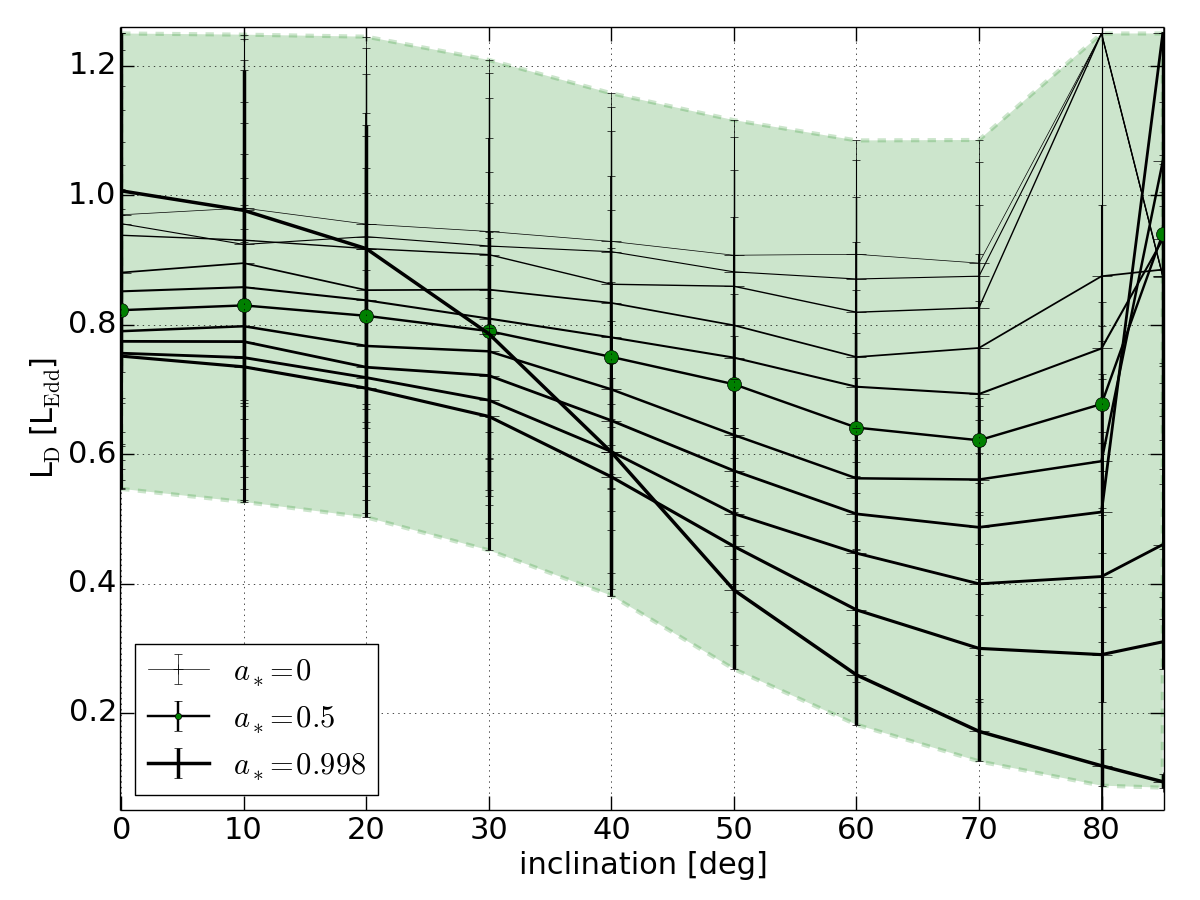}
 \caption{The disc luminosity in HLX-1. {\it Left}: The shaded areas indicate the 90\% confidence interval for the luminosity distribution associated with the black hole mass measured in Figure~\ref{fig:mass}. Again, the the result for $a_* = 0.5$ is singled out and given by the three solid lines (with 90\% confidence), corresponding, respectively, to the {\it Chandra} (grey), {\it Swift} (green) and {\it XMM} (blue) spectrum. {\it Right}: The influence of the black hole spin on the luminosity of the {\it Swift} spectrum. Error bars give the 90\% confidence interval.}
 \label{fig:lumi}
 \end{figure*}

The {\it XMM} spectrum requires a significant amount of Comptonisation to model the high energy tail (see Figure~\ref{fig:spectra}). The contribution of the disc to the total luminosity is thus substantially lower than for the {\it Swift} and {\it Chandra} spectrum, as shown in the left panel of Figure~\ref{fig:lumi}. Both increasing spin and increasing inclination reduce the effective emission area which is efficiently compensated with increasing black hole masses during the fit procedure, while the disc luminosity does not change significantly. If, however, the configuration is set to maximum spin, the model spectrum becomes so soft below the peak that tuning the mass parameter alone is not sufficient anymore to fit the observed spectrum, instead the disc luminosity needs to be increased. This behaviour is visible in the right panel of Figure~\ref{fig:lumi}, where the luminosity that corresponds to the maximum spin crosses all other lines. In practice this means that in a low inclination system with uncertain black hole mass a low spin could be easily mistaken for maximally rotating black hole.
% ================================================
% FIGURE -- 4
% ================================================
% ================================================

There is currently no evidence of eclipses in HLX-1, the inclination is thus likely to be lower than about $75^{\circ}$. We assume here an inclination of $60^{\circ}$ to demonstrate the modelling of the three spectra for a moderate spin, $a_* = 0.5$, in Figure~\ref{fig:spectra} and the resulting parameter values in Table~\ref{tab:results}. Because of the weakness of the high energy tail in the {\it Swift} and {\it Chandra} spectra the photon index, $\Gamma$, is only weakly constrained. Moreover, in the chosen example the adding of a Compton component is statistically not very significant, in particular for the {\it Chandra} spectrum; the F-test in XSPEC gives a probability of $p = 3 \times 10^{-18}$, $p = 0.008$, and $p = 0.178$ for the {\it XMM}, {\it Swift}, and the {\it Chandra} data, respectively. However, as we move through the parameter space to lower black hole spins and inclinations, the Compton component becomes more significant, even for the {\it Swift} and {\it Chandra} observation. This is because for low spin and low inclination pure absorbed disc models are intrinsically too soft to properly reproduce the high energy (> 4keV) photons. As a consequence, the models with the lowest spins and inclinations, $a_* \lesssim 0.1$ and $i \lesssim 20^{\circ}$, respectively, favour high disc luminosities; without the Compton component they would require a luminosity higher than our current limit of $L = 1.25 \LEdd$. Whether such models would ultimately fit the data better remains to be seen in future work.

\subsection{The mass accretion rate}
\label{sec:efficiency}
\begin{figure}[ht]
 \includegraphics[width=0.5\textwidth]{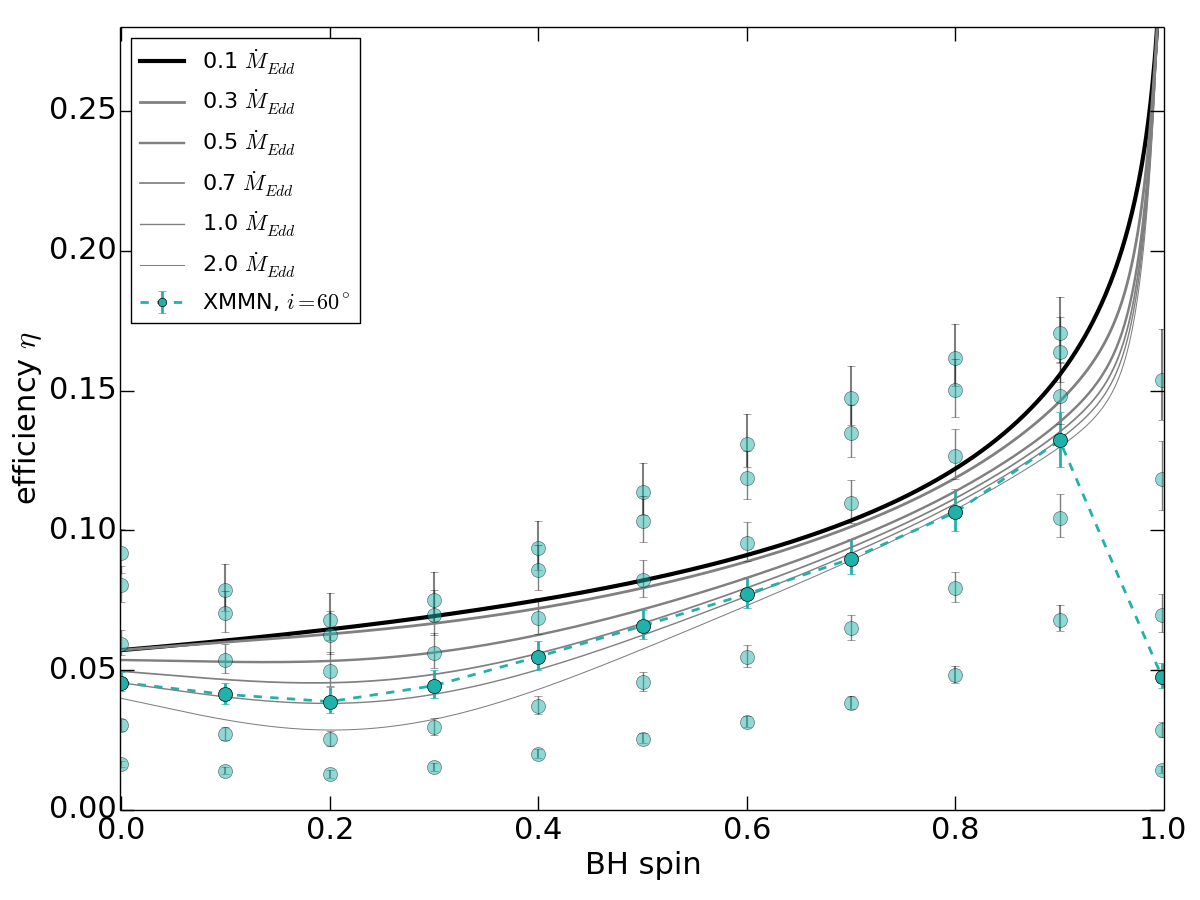}
 \caption{Slim disc efficiency. The solid lines indicate model efficiencies for a number of mass accretion rates. The black line corresponds to the standard thin disc efficiency. For HLX-1 we derive effective efficiencies from the fitted disc luminosity of the {\it XMM-Newton} spectrum and the observed unabsorbed 0.3-10keV disc luminosity for selected inclinations (i = $0^{\circ}$, $30^{\circ}$, $50^{\circ}$, $60^{\circ}$, $70^{\circ}$, and $80^{\circ}$). The dashed line highlights the spin evolution of the effective efficiency for a disc with i$= 60^{\circ}$. Error bars include the uncertainty in black hole mass and disc luminosity. }
 \label{fig:eff}
 \end{figure}

The efficiency of an accretion disc, $\eta = 1 - u_t(r_{\rm in})/c^2$, defines how much of the accreted mass can be converted into luminosity, 
\begin{equation}
L = \eta \dot{M} c^2 \, .
\label{eq:L}
\end{equation} 
The relevant quantity is the specific energy at the inner disc edge, $u_t(r_{\rm in})$. While the inner disc radius in standard discs is given by the ISCO, so that the standard disc efficiency, $\eta_*$, only depends on the black hole spin, in slim discs there is an additional dependency on the mass accretion rate that moves the slim edge from the ISCO towards the marginally bound radius, $r_{\rm mb}$. This means the slim disc efficiency decreases for increasing mass accretion rates and formally goes to zero because $u_t(r_{\rm mb}) = 1$. In Figure~\ref{fig:eff} we show the behaviour of the standard and slim disc efficiency with black hole spin. To obtain the mass accretion rate from the fitted model luminosity we first rewrite Equation~(\ref{eq:L}) using the normalised mass accretion rate $\dot{m} \equiv \Mdot / \MdotEdd$ and $\MdotEdd  \equiv \LEdd /(\eta_*  c^2)$ for the Eddington limit on the mass accretion rate\footnote{Another common definition is $\MdotEdd  = \LEdd / c^2$.}. Then we  employ the approximate formula of the potential spout inner edge given in \citet{abr+10},
\begin{eqnarray}
r_{\rm in} = & {\rm Min} & \left[ (0.275 - 0.410 \, a_* + 0.143 \, a_*^2) \, \dot{m}^{-1.4} \, + \right. \nonumber\\
                       &&  \left. 4.45  4.87 \, a_* + 8.06 \, a_*^2 - 6.38 \, a_*^3 \; \right. \\
                       && \left. 0.985 \, r_{\rm ISCO} \right]  \, , \nonumber 
\end{eqnarray} 
to find the inner radius of the slim disc and obtain an expression for $\eta$ that depends only on $a_*$ and $\dot{m}$. From the slim disc fits one obtains the modelled disc luminosity, so that Equation~(\ref{eq:L}) finally can be solved for $\dot{m}$. This results in mass accretion rates between $\Mdot \simeq 7 \times 10^{-5}$ and $2 \times 10^{-3} \, \Msun/{\rm y}$.  The highest mass accretion rates are achieved for slowly rotating black holes with highly inclined accretion discs. This agrees well with the results of \citet{god+12} who calculated the mass accretion rate during the outbursts to be $\sim 1.2 \times 10^{-4} \, \Msun/{\rm y}$ and $\sim 8.2 \times 10^{-5} \, \Msun/{\rm y}$ assuming a Schwarzschild black hole and a face-on disc. The corresponding model efficiency takes values between $\eta \simeq (0.045 - 0.049) \pm 0.001$ for a Schwarzschild black hole (with i = $85^{\circ} - 0^{\circ}$) and $\eta \simeq 0.32$ for a maximally rotating Kerr black hole (independent of inclination). Next, we look at the observed bolometric disc luminosity, $L_{\rm bol} = 4 \pi D^2 F_{\rm obs}$, calculated from the 0.3-10keV unabsorbed disc flux. One can use Equation~(\ref{eq:L}) to compare the observed luminosity output to the fitted mass accretion rate and obtain the \emph{effective efficiency} of the accretion disc in HLX-1. We show the results in Figure~\ref{fig:eff}. As \citet{god+12} state, a Schwarzschild black hole with a face-on disc is relatively efficient with $\eta \simeq 0.1$. This also holds for low inclinations, i < $30^{\circ}$, and moderate black hole spins, $0.4 < a* < 0.9$. We find that starting from an inclination of $30^{\circ}$ the disc becomes increasingly inefficient, in particular for minimal and maximal spins. Above an inclination of $\sim 50^{\circ}$ the disc has become an entirely inefficient emitter, i.e. the effective efficiency is for all spins significantly lower compared to a standard disc. This means on the one hand that the best fitting models to inclined discs require substantial mass accretion rates that can only be matched to the comparatively low luminosity output if the disc radiates inefficiently. On the other hand, if we see the disc nearly face-on, even a small accretion rate may generate a huge luminosity. To determine whether the disc in HLX-1 is advection dominated or over-efficient, better constraints on the inclination are required. 

% ================================================
% FIGURE -- 5
% ================================================
% ================================================

\subsection{The effect of viscosity}
\label{sec:viscosity}
In hydrodynamic disc models like {\tt bhspec} and {\tt slimbh} the relation between viscous stresses and pressure is parametrised by the viscosity parameter, $\alpha$. Although accretion discs have been studied for forty years, the precise nature of this relation is still unclear and $\alpha$ remains a phenomenological quantity that possibly comprises several physical mechanisms related to the pressure balance in accretion discs. We use a default value of $\alpha = 0.01$ which is supported by observations of high luminosity, radiation pressure dominated discs that require a roughly constant colour temperature correction (i.e. hardening factor) which is only supported by a low viscosity \citep{don+08}. We note, however, that outburst cycles in low mass X-ray binaries require a very efficient angular momentum transport and thus suggest a high $\alpha$ value \citep[see e.g. the review by][]{las01}. We therefore perform an additional  analysis assuming $\alpha = 0.1$ and find that the black hole mass increases by 7\%-14\%. Figure~\ref{fig:spectra} shows the corresponding best fit as dashed lines. Since a larger viscosity translates into a smaller amount of thermalised photons in the plasma, which in turn entails a larger colour temperature correction, the slim disc spectra of a given luminosity get harder with increasing $\alpha$. This change in the spectral shape causes a roughly 10\% increase in mass. We caution, however, that a high luminosity disc becomes effectively optically thin when the viscosity is too large and therefore consider the low viscosity value as the standard \citep[see also the discussion in][]{str+11, str+13}.

% ================================================
% ==============================================
\section{Conclusions}
\label{sec:discussion}
% ================================================
% ================================================
We analysed three spectra of the IMBH candidate HLX-1 that were collected by {\it Swift}, {\it XMM-Newton}, and {\it Chandra} during different missions between 2008 and 2012. We estimate the black hole mass using the fully relativistic slim disc model, {\tt slimbh} \citep{sad+11, str+11}, which allows us to self-consistently probe the trans-Eddington luminosity regime in the whole parameter plane spanned by black hole spin and inclination. This addresses and remedies the deficits of previously used models which were either not relativistic \citep{ser+11}, only valid at lowest luminosities \citep{dav+11}, or only valid for one particular inclination and spin \citep{god+12}.  Assuming a low disc viscosity ($\alpha = 0.01$) we find that a Schwarzschild black hole has a mass of about 6,300 - 50,900 $\Msun$ (increasing with disc inclination), whereas a maximally spinning black hole has a mass between 16,900 - 191,700 $\Msun$. A high viscosity disc ($\alpha =0.1$) has black hole masses that are roughly 10\% higher.  This result is consistent among all three observations with {\it Swift}, {\it XMM-Newton}, and {\it Chandra}. Moreover, it is also in good agreement with earlier measurements based on {\tt bhspec} \citep{dav+11} and other slim disc models \citep{god+12}. 

The continuum fitting method that we have applied here determines the inner edge of the accretion disc given its effective temperature and flux, i.e. it is designed to measure the black hole spin and relies strongly on the knowledge of the binary parameters, M , D, and i. Given that the inclination is only constrained to $i < 75^{\circ}$ the constraints on the black hole mass are necessarily fairly weak. Nonetheless, our results clearly place HLX-1 in the regime of intermediate mass black holes. Future dynamical measurements of the binary parameters of HLX-1 will allow us to apply the continuum fitting method as it has been intended, namely to assess the spin of the IMBH.

% ================================================
% ================================================
\section*{Acknowledgments}
% ================================================
% ================================================
We thank S. Farrell for his invaluable feedback on the original manuscript and the anonymous referee for their helpful remarks. OS thanks in particular M. Bursa for the continuous and illuminating discussions about slim discs.

\bibliographystyle{aa}
\bibliography{slimbh_mass_of_HLX1}

\end{document}